\documentclass{svjour3}
\usepackage{latexsym}
\usepackage{amssymb}
\usepackage{graphicx,epstopdf}

\usepackage{color}

\newcommand{\bla}{\color{black}}

\newcommand{\half}{\frac{1}{2}}
\newcommand{\sep}{\textrm{sep}}
\date{\today}
\begin{document} 
\title{Toward secure communication using intra-particle entanglement}
\author{S. Adhikari \and Dipankar Home \\ A. S. Majumdar \and
A. K. Pan \\ Akshata Shenoy H. \and R. Srikanth}

\institute{S. Adhikari \at 
	Indian Institute of Technology, Jodhpur, India \\
\email{tapisatya@gmail.com}
\and
Dipankar Home \at
CAPSS, Dept. of Physics, Bose Institute, Salt Lake,
Kolkata-700091, India \\
\email{dhome@bosemain.boseinst.ac.in}
\and
A. S. Majumdar \at 
S. N. Bose National Centre for Basic Sciences,
Salt Lake, Kolkata 700 098, India \\
\email{archan@bose.res.in}
\and
A. K. Pan \at
Dept. of Physics, NIT Patna, India\\
\email{alokkrpan@gmail.com}
\and
Akshata Shenoy H. \at
Electrical Communication Engineering Dept., IISc, Bangalore, India
\email{akshata@ece.iisc.ernet.in}
\and
R. Srikanth \at
Poornaprajna Institute of Scientific Research, Bangalore, India
\email{srik@poornaprajna.org}
}

\date{Received: date / Accepted: date}

\maketitle

\begin{abstract}
We explore the use of  the resource of intra-particle entanglement for
secure  quantum key distribution  in the  device-independent scenario.
By virtue of the local nature of such entanglement, Bell tests must be
implemented  locally, which leads  to a  natural decoupling  of device
errors from channel errors.  We  consider a side-channel attack on the
sender's     state    preparation     device,     for    which     the
intra-particle entanglement-based  scheme is  shown to be  more secure
than the one that uses separable states. Of practical relevance is the
fact  that such  entanglement  can be  easily  generated using  linear
optics.
\end{abstract}


\section{Introduction}

Quantum  key  distribution  (QKD)  protocols  \cite{gisin}  allow  two
distant  parties, traditionally  called Alice  and Bob,  to  produce a
shared random bit string consisting of 0's and 1's known only to them,
which can be used as a  key to encrypt and decrypt messages.  Based on
fundamental  principles  such  as  the  quantum  no-cloning  principle
\cite{wootters},  QKD  provides   an  unconditionally  secure  way  to
distribute random  keys through insecure  channels. 

While the first QKD scheme to be proposed, the Bennett \textendash Brassard (BB84)
protocol \cite{bb84}  was a prepare-and-measure  protocol, which used
separable  states, a  connection between  nonlocality \cite{bell,chsh}
and security  was first suggested by the  Ekert protocol \cite{ekert}.
The basic  intuition here is that  Eve's attack causes  a reduction in
the correlation between legitimate parties, which is now understood as
due  to   the  monogamy  of  nonlocality   in  non-signaling  theories
\cite{mag06}.  It  is also known  that nonlocality helps  security not
only in the  traditional QKD scenario (where Eve  attacks the channel)
but even in the more stringent device-independent (DI) scenario, where
neither  the prepared  initial  states nor  the  devices are  trusted.
Security  here  must  be  guaranteed simply  via  certain  statistical
checks \textemdash typically sufficiently high  violation of a Bell inequality \textemdash
and   without  requiring  a   detailed  characterization   of  devices
\cite{MY98,BHK05,SGBMPA06,MPA11,pir09,VV12}.

The   nonlocality,  and  hence   entanglement,  considered   above  is
\textit{inter}-particle   entanglement,    and   the   Bell-inequality
violating  property (i.e., nonlocality)  pertains to  the correlations
obtained by  spatially separated measurements by a  sender (Alice) and
receiver (Bob).  A different kind  of entanglement is that between two
degrees of freedom of the \textit{same} particle, i.e., intra-particle
entanglement.    This   has   been    discussed   by   Basu   et   al.
\cite{Home-Kar-PLA-2001}  in  the   context  of  a  Mach \textendash Zehnder  type
interferometric   set-up   for    demonstrating   the   violation   of
non-contextuality.  An experiment  using single neutrons was performed
by Hasegawa et al.   \cite{hasegawa-nature-2003}.  

Here,  for the  first time,  we  propose the  use of  the resource  of
intra-particle entanglement for  QKD.  By its nature,  Bell tests with
intra-particle entanglement must be  local.  Interestingly, such local
Bell tests  have also been proposed  in the case of  bipartite systems
for  self-testing schemes  employed to  certify the  state preparation
process or the source  of quantum states \cite{LPT+12,MY04,TH11}.  Our
method also evokes a comparison with one-sided DIQKD,
in  which the  Alice's devices  are untrusted,  but Bob's  are trusted
\cite{BCW+12},  and  where   the  statistical  check is  based  on
steering inequalities  \cite{WJD07}.  An  interesting counterpoint
here is provided  by the scenario of  measurement DI,
in which, the devices for measurement, rather than that of the sender,
is untrusted \cite{LCQ12}.

Relative to  inter-particle entanglement,  intra-particle entanglement
is easy  to generate.   In the optical  case, considered  here, linear
optics  suffices.    However,  the  local  nature   of  intra-particle
entanglement means that it is  unsuitable for many quantum information
processing tasks, like  quantum teleportation or dense  coding.  It is
an interesting question  whether it is useful  for cryptography, which
we  answer here  in the  affirmative.  Experimental  demonstrations of
various QKD protocols were discussed in 
\cite{lucamarini,bruss,wang,kraus,lo,adachi}.  
The practical violation of the Bell's inequality
in the cryptographic context was  first considered in an experiment by
Jennewein et.   al. \cite{jennewein},  but no quantitative  measure of
security was derived from the observed violation. Later, Ling et.  al.
\cite{ling}  performed an  experiment  on entanglement-based QKD,  in
which the violation  of Bell-CHSH inequality is used  to also quantify
the  degree   of  security  according   to  the  criterion   of  Refs.
\cite{mag06,AGM06}.

In  this work,  our  accent is  mainly  on introducing  intra-particle
entanglement between position (path) and polarization of photons, as a
useful  and easy-to-prepare  resource  for QKD,  which presents  novel
elements when state  preparation devices, in addition  to the channel,
are allowed  to be  insecure. We  do so  by showing  that this  QKD is
secure against certain  ``side channels'' that leak  secret data (such
as Alice's  or Bob's  settings and  outcome information),  whereas the
corresponding version  of BB84  is not secure.   Clearly, there  is no
protection against an unrestrictedly  powerful side channel.  Hence, we
must assume that  it cannot be ``obvious''.  Examples  of typical side
channels are timing  information on the devices  used, observations of
power consumption or electromagnetic leaks bearing some heat signature
of devices, or even a click  sound produced by an optical element.  We
quantitatively find that the  Bell-inequality violating (BIV) property
of  the  path-polarization  correlations  can  be  used  to  guarantee
security against an individual side-channel attack, which would render
insecure QKD in the standard  scenario.  Here an incoherent attack
is  one where  Eve attacks  Alice's particles  along the  transmission
channel  individually and  measures  them  independently, without  the
involvement of  any joint  measurement.  Further details,  such as
coherent  attacks,  optical  losses   in  the  channel  and  universal
composability  are  important  future  directions of  this  work,  not
considered here.   An important  aspect of using  intra-particle-based
entanglement, which we consider  in more detail elsewhere \cite{Ax++},
is to generalize the  Goldenberg \textendash Vaidman protocol for orthogonal state-
based  cryptography  \cite{GV95},  as   a  method  to  thwart  general
individual  and  coherent   attacks  \cite{CBK+02}  on  intra-particle
entanglement-based qudits.

This  article is  divided  as follows:  in  Sect. \ref{sec:ipe},  we
introduce  the  notion  of  intra-particle entanglement and  present
simple   generation  schemes   for   path-polarization  intra-particle
entanglement.   In Sect. \ref{sec:scha},  we introduce  an augmented
key  distribution   protocol,  suitable  for   a  side-channel  attack
scenario,  in  which the  sender  Alice  must  verify the  quality  of
intra-particle entanglement  just before transmitting  the particle to
the receiver Bob, and after all optical elements used for the encoding
process have  been applied.  In Sect.  \ref{sec:cheq}, an individual
attack scheme by  Eve is considered.  Her action  is to depolarize the
initial maximally entangled state into an intra-particle Werner state,
for which the entanglement and  BIV properties are readily known.  The
condition  for  secure  extraction   of  secret  bits  is  studied  in
comparison  with the  availability of  these properties  in  the noisy
state received  by Bob.  In Sect.  \ref{sec:sidecha}, we demonstrate
the usefulness of intra-particle  entanglement in protecting against a
class of  side-channel attacks that  rely on flaws in  certain optical
elements  such  as  quarter-wave  plates (QWPs).   Finally,  we  present  our
conclusions in Sect. \ref{sec:conclu}.

\section{Intra-particle entanglement \label{sec:ipe}}

\begin{figure}[!ht]
\includegraphics[width=10.0cm]{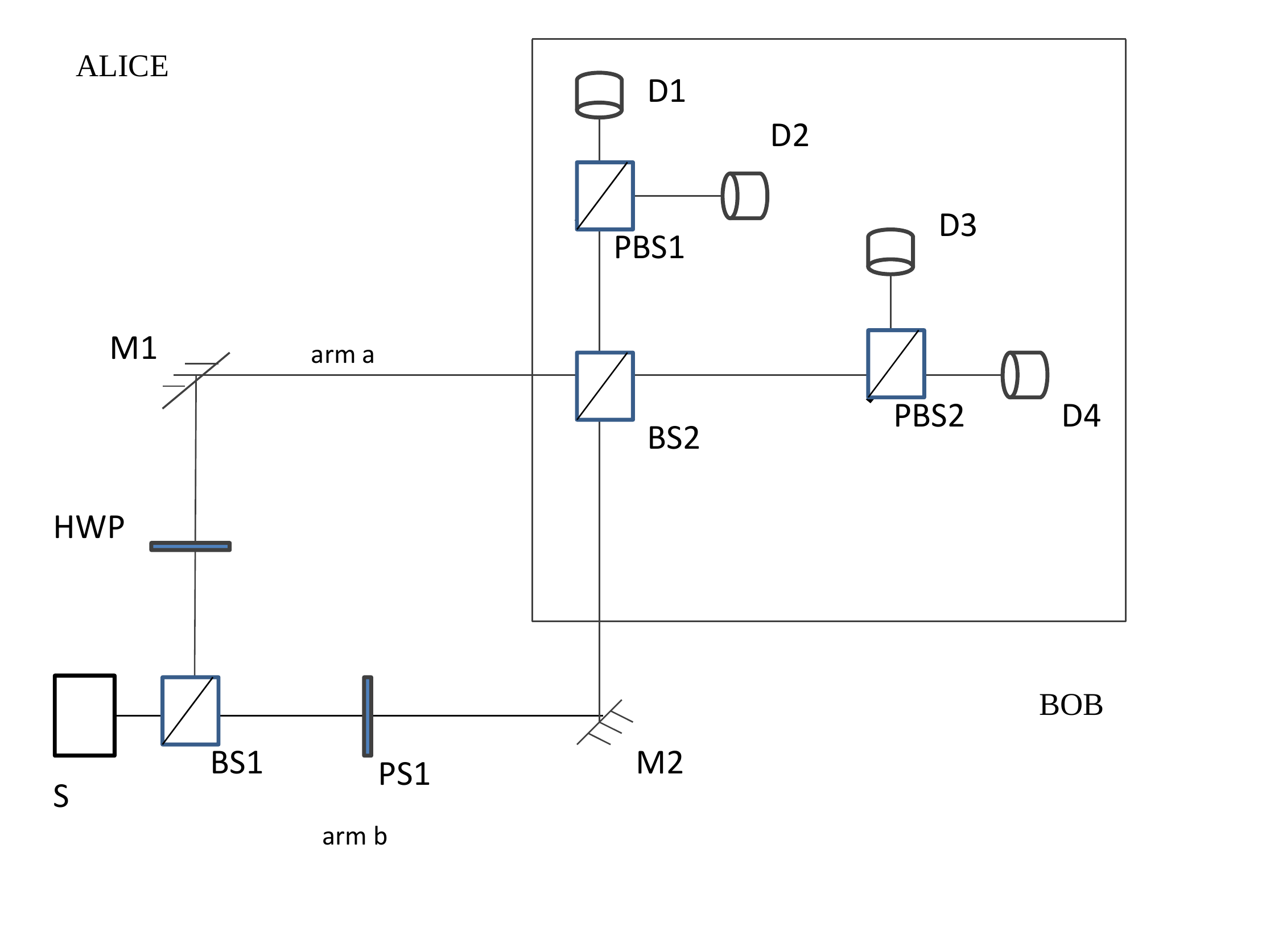}
\caption{BB84 set-up:  Alice transmits  a state to  Bob in one  of the
  bases $G_j^A$  by suitably applying  the linear optical  elements of
  beam splitters, HWP, QWP and PS. Bob may recombine the reflected and
  the  transmitted channels at  BS2.  Finally,  Bob performs  path and
  polarization measurements  using the polarizing  beam splitters PBS1
  and PBS2.}
\label{fig:bb84}
\end{figure}

Let  us  consider a  photon  that  is  initially polarized  along  the
vertical direction  (its state  denoted by $|0\rangle$).   Taking into
consideration   its   path   (or   position)  variables,   the   joint
path-polarization state can be written as
\begin{equation}
|\psi_{0}\rangle_{ps}=|V\rangle_{s} \otimes |\psi_{0}\rangle_{p}
\label{pathspinpro}
\end{equation}
where the subscripts $p$ and $s$ refer to the path and the spin (i.e.,
polarization)   variables,  respectively.   A   photon  in   the  state
$|\psi_{0}\rangle_{ps}$  with Alice  is incident  on a  beam splitter
(BS1),   whose   transmission   and   reflection   probabilities   are
$|\alpha|^{2}$      and     $|\beta|^{2}$      respectively,     where
$|\alpha|^{2}+|\beta|^{2}=1$ (cf. Fig. \ref{fig:bb84}).

The  reflected  and transmitted  states  from  BS1  are designated  by
$|\psi_{R}\rangle$  and  $|\psi_{T}\rangle$,  respectively.   Here  we
recall that for any given  lossless beam splitter, arguments using the
unitarity condition show  that for the particles incident  on the beam
splitter, the  phase between the transmitted and  the reflected states
of the particle  is $\frac{\pi}{2}$. Note that the  beam splitter acts
only on  the path states without  affecting the polarization  state of
the particles.

The state of a particle emergent from BS1 can then be written as
\begin{eqnarray}
|\psi_{0}\rangle_{ps} \rightarrow   |\psi_{1}\rangle_{ps}  =  |V\rangle_s
\otimes (\alpha|\psi_{T}\rangle_{p}+i\beta|\psi_{R}\rangle_{p}),
 \label{pathspinpro1}
\end{eqnarray}
where 
\begin{eqnarray}
&&|\psi_{T}\rangle_{p}\equiv \left(%
\begin{array}{c}
  0   \\
  1  \\
  \end{array}%
\right),|\psi_{R}\rangle_{p}\equiv \left(%
\begin{array}{c}
  1   \\
  0  \\
  \end{array}%
\right) {}\nonumber\\&&
|V\rangle_{s}\equiv |0\rangle_{s}\equiv \left(%
\begin{array}{c}
  0   \\
  1  \\
  \end{array}%
\right),|H\rangle_{s}\equiv |1\rangle_{s}\equiv \left(%
\begin{array}{c}
  1   \\
  0  \\
  \end{array}
\right). \label{notations}
\end{eqnarray}
Our simplest basis, called  $G_1^A$, can be generated without using
the beam splitter:
\begin{eqnarray}
|\Psi_+\rangle &=& |0\rangle_{s}\otimes|\psi_{T}\rangle_{p},
\nonumber\\
  |\Psi_-\rangle &=& |1\rangle_{s}\otimes|\psi_{T}\rangle_{p},
\nonumber\\
|\Psi^\ast_+\rangle &=& |0\rangle_{s} \otimes |\psi_{R}\rangle_{p},
\nonumber\\
  |\Psi^\ast_-\rangle &=& |1\rangle_{s}\otimes|\psi_{R}\rangle_{p}.
\label{fourstate0}
\end{eqnarray}

A basis consisting of  path-polarization entangled elements and which
is  mutually unbiased  with $G_1^A$,  is $G_2^A$,  given below  in Eq.
(\ref{fourstate}).   It  is  produced   by  a  linear  optical  set-up
consisting of a beam splitter, a half-wave plate (HWP), QWP
  and a phase shifter (PS).   For example, $|\Phi_+\rangle$
is produced  from $|\Psi_+\rangle$, by passing the  particle through a
BS1,    applying    HWP     on    the    transmitted    wave    packet
$|\psi_{T}\rangle_{p}$,  followed by  the application  of QWP  on both
arms.   The  HWP   has  the   action   $|H\rangle_{s}  \leftrightarrow
|V\rangle_{s}$. The  states resulting  from $G_1^A$ by  this procedure
are:
\begin{eqnarray}
|\Phi_{\pm} \rangle &=& \frac{1}{\sqrt{2}}\left(\frac{|0\rangle_{s}
 + |1\rangle_{s}}{\sqrt{2}}
  \otimes|\psi_{T}\rangle_{p} \pm i
  \frac{|0\rangle_{s}-|1\rangle_{s}}{\sqrt{2}}\otimes|\psi_{R}\rangle_{p}\right),\nonumber\\
  |\Phi^\ast_\pm\rangle &=& \frac{1}{\sqrt{2}}\left(\frac{|0\rangle_{s}-|1\rangle_{s}}{\sqrt{2}}
  \otimes|\psi_{T}\rangle_{p} \pm i
  \frac{|0\rangle_{s}+|1\rangle_{s}}{\sqrt{2}}\otimes|\psi_{R}\rangle_{p}\right),
\label{fourstate}
\end{eqnarray}
which  form the  basis $G_2^A$.   The bases  $G_1^A$ and  $G_2^A$ are
mutually unbiased in the sense that  any element in either basis is an
equal weight  superposition (apart from phase factors)  of elements of
the other basis.  

To   measure    the   state   in   an    arbitrary   separable   basis
$(\vec{a},\vec{b})     \equiv     \vec{a}\cdot\vec{\sigma}     \otimes
\vec{b}\cdot\vec{\sigma}$,  where   $\vec{a}$  and   $\vec{b}$  denote
direction vectors of unit magnitude, one passes the particle through a
beam splitter  of suitable bias that de-rotates the position  to the
computational basis,  and then uses  two detectors, both set  alike to
measure the  polarization along  $\vec{a}$.  For example,  if $\vec{b}
\equiv  (\sin\theta\cos\phi,  \sin\theta,\sin\phi,  \cos\theta)$,  the
beam  splitter  is  chosen   with  coefficients  of  transmission  and
reflection   being  $\cos(\theta/2)$   and  $e^{i\phi}\sin(\theta/2)$,
respectively.   These  coefficients can  be  set  at the  time  of
manufacture by the reflection coating  applied to a beam splitter, and
their ratio  determines whether  the element  functions as  a balanced
(50:50)  or unbalanced  (say 90:10) beam  splitter.  The  particular
measurements basis  settings $(\vec{a}, \vec{b})$ we  require in order
to    evaluate   the    Bell    observable,   are    given   in    Eq.
(\ref{eq:settings}). 

In dimension $d=4$, there  are $d+1=5$ mutually unbiased bases (MUBs).
Another  mutually  unbiased   entangled  basis,  denoted  $G_3^A$,  in
addition to sets $G_1^A$ and $G_2^A$, is:
\begin{eqnarray}
|\Lambda_{\pm} \rangle &=& \frac{1}{\sqrt{2}}\left(\frac{|0\rangle_{s}
 + i|1\rangle_{s}}{\sqrt{2}}
  \otimes|\psi_{T}\rangle_{p} \pm i
  \frac{|0\rangle_{s}-i|1\rangle_{s}}{\sqrt{2}}\otimes|\psi_{R}\rangle_{p}\right), \nonumber\\
  |\Lambda^\ast_\pm\rangle &=& \frac{1}{\sqrt{2}}\left(\frac{|0\rangle_{s}-i|1\rangle_{s}}{\sqrt{2}}
  \otimes|\psi_{T}\rangle_{p} \pm i
  \frac{|0\rangle_{s}+i|1\rangle_{s}}{\sqrt{2}}\otimes|\psi_{R}\rangle_{p}\right).
\label{fourstate1}
\end{eqnarray}
Two others  (separable state) MUBs,  which may be denoted  $G_4^A$ and
$G_5^A$,  can be  produced by  applying  $H \otimes  H$ and  $H^\prime
\otimes   H^\prime$  to   the   elements   of  basis   $G_1^A$
(\ref{fourstate0}), where $H \equiv \frac{1}{2}(\sigma_z + \sigma_x)$,
while $H^\prime  \equiv \frac{1}{2}(\sigma_z + \sigma_y)$.   All these
states are easy to prepare, requiring only linear optical elements.

Alice's  states are  analyzed in  Bob's system,  consisting of  a beam
splitter (BS2), followed by  polarization analyzer in each output arm.
For   example,   if   she   sends  the   state   $|\Phi_+\rangle$   or
$|\Phi_{-}\rangle$,  then   after  emerging  from   BS2  (cf.   Fig.
\ref{fig:bb84}), the corresponding resulting  states at Bob's site are
given by
\begin{eqnarray}
  |\Phi^\prime_+\rangle&=&\frac{1}{\sqrt{2}}(i|\chi_{1}\rangle\otimes|\psi_{T}'\rangle+
  |\chi_{2}\rangle\otimes|\psi_{R}'\rangle){},\nonumber\\
  |\Phi^\prime_{-}\rangle&=&\frac{1}{\sqrt{2}}(i|\chi_{2}\rangle\otimes|\psi_{T}'\rangle+
  |\chi_{1}\rangle\otimes|\psi_{R}'\rangle),
 \label{beamsplitter2}
\end{eqnarray}
where
$|\chi_{1}\rangle=|0\rangle_{s}$,     $|\chi_{2}\rangle=|1\rangle_{s}$,
$|\psi_T\rangle   =   \frac{1}{\sqrt{2}}\left(|\psi^\prime_T\rangle   +
i|\psi_R^\prime\rangle\right)$       and       $|\psi_T\rangle       =
\frac{1}{\sqrt{2}}\left(|\psi^\prime_T\rangle                          -
i|\psi_R^\prime\rangle\right)$.

\section{Protocol \label
{sec:scha}}

In  each  of  the  five  MUBs,  given  by  states  (\ref{fourstate0}),
(\ref{fourstate}), (\ref{fourstate1}),  etc., Alice and  Bob designate
basis  elements by  numbers 0,  1,  2 and  3.  We  can form  different
protocols by  considering any  two or  more of the five bases.   It will
suffice for us to consider the protocol $\mathcal{P}_{1,2}$ where only
two bases are used, which  are $G_1^A$ and $G_2^A$. However, any other
pair of  MUBs will  do. Using  a larger number  of bases  enhances the
tolerable error  rate, but is  less efficient and  experimentally more
difficult.

The protocol is as follows. (1) Using mirrors, phase shifters and beam
splitters  (Fig.  \ref{fig:bb84}),  Alice  prepares an  intra-particle
path-polarization  entangled  states  in  basis  $G_1^A$  or  $G_2^A$,
randomly   chosen,    starting   from   the   initial    input   state
$|V\rangle|\psi_T\rangle$.  (2)  Just before transmission to  Bob, but
after state preparation, she selects a  fraction $g$ of the states, to
verify that  their fidelity with  the intended output  entangled state
remains 1.  She  may do so by reversing her  preparation procedure and
observing  the probability  $e_A$  that the  output \textit{fails}  to
coincide with  $|V\rangle|\psi_T\rangle$. As explained later,  she may
equivalently perform a test of Bell inequality violation. This step is
a key addition to the protocol for protection against the side-channel
attack;  Here we  assume that  the  optical set-up  that reverses  the
preparation   is  different   from  that   actually  used   for  state
preparation, though this is not necessary;   we note that the test
based  on reversing  implicitly  verifies that  the  reversed state  is
separable. Since an entangled bipartite  state is necessarily mixed in
any  one  of  its  parts,  a  test  of  mixedness  may  be  optionally
implemented. If it returns a  positive result, then clearly the output
fails to  coincide with $|V\rangle|\psi_T\rangle$.  (For  a particular
proposal for  testing mixedness,  see Ref. \cite{MPA13}).  (3)
Alice  transmits the  remaining  particles to  Bob  (i.e., no  quantum
memory is  used to  hold the  other particles while  the Bell  test is
underway);  (4)  Bob  obtains  a   2-bit  outcome  by  using  mirrors,
phase shifters and  beam splitters  (Fig.  \ref{fig:bb84})  to measure
the  transmitted  photon,  choosing  randomly  the  measurement  basis
$G_j^B$ basis, the "primed" versions of $G_j^A$.

The remaining  steps involve  only classical post-processing:  (5) The
experiment described  in the above  two steps is repeated  many times.
Alice then  declares via an authenticated classical  channel the value
of $e_A$  and the  basis (but  not the basis  element) from  which she
chose the  state  (The existence of an  authenticated channel between
Alice and Bob,  which gives Bob an edge over Eve,  is essential to the
security  of  QKD).  Bob  announces  the cases  where  his  basis  was
mismatched with  hers.  The corresponding measurement  outcome data are
discarded.   (6)  From  the  retained  (sifted)  measurement  data,  a
sufficiently large  portion is  divulged by Bob.   The fraction  of it
that does not  agree with her preparation state is  an estimate of the
error rate in the key, $e$.   If $e$ is sufficiently low, they proceed
with the rest of the protocol,  else they abort it.  (7) Alice and Bob
perform key  reconciliation over  the authenticed channel,  to improve
the  correlation of  their  respective  copy of  the  key.  They  then
perform  privacy amplification  to minimize  Eve's information  on the
key.

The verification step  in the protocol, which is  an augmentation over
conventional QKD  protocols, makes  our protocol more  secure in  a DI
scenario,  as  we explain  later.   With inter-particle  entanglement,
Alice  and  Bob  perform  this  step by  using  local  operations  and
classical  communication (LOCC)  to determine  the nonlocality  of the
state, from  which an estimate  of the secrecy content  follows.  With
intra-particle  entanglement, only  one of  the parties  (here, Alice)
must accomplish this, because  the path and polarization qubits cannot
be measured at spatially separated stations.  In this sense, Alice and
Bob  distinguish  between  errors  arising  due to  the  channel  (the
conventional  security scenario)  versus errors  arising  during state
preparation (the side-channel or DI) scenario.

Before discussing detailed security issues, let us make a number of
remarks to qualify the motivation behind this work. 
\begin{enumerate}
\item  Formally, the four-dimensional  spin-position entangled  states we
  use for  encoding can be  considered as superpositions of  a ququart
  (four-dimensional quantum  system). However, our point is  to study the
  system as two-qubit entanglement because  we wish to take advantage of
  some ideas from DI quantum cryptography.
\item  We   consider  below  (apart  from  a   channel  attack)  Eve's
  side-channel  attack only  on  Alice's QWP,  and  not other  optical
  elements. There is  no special reason to make  this choice.  Rather,
  this is  meant only as an  illustration of the  general principle of
  how  intra-particle entanglement  can be  more useful  that  a plain
  qubit  or  ququart superposition.  An  more  detailed follow-up  can
  extend this principle to other optical elements.
\item  Note that the  verification step  itself can  be subject  to an
  attack,  for  example,  to  the particular  side-channel  attack,  we
  describe below.  What this entails is  that Eve can  know what Alice
  knows  about  Eve's   eavesdropping  actions,  possibly  introducing
  further noise.  From the viewpoint of enhancing her knowledge of the
  final  key, this  does not  help Eve,  and we  do not  consider such
  attacks here.
\end{enumerate}

\section{Security in the conventional scenario \label{sec:cheq}}

We model  Eve's attack as  a simple intercept-resend attack  on single
particles.   Eve  can  also  eavesdrop their  authenticated  classical
channel  and  thus make  use  of  their  basis announcements.   Eve's
strategy is to measure the particles randomly in one of the legitimate
bases.  She forwards the measured  state to Bob, and waits until after
their public announcement of bases to  find out when she got it right.
For  purposes of  this  section, Eve  is  assumed to  attack only  the
channel  and   not  exploit   any  side  channels.   Accordingly,  the
verification step (2) in the protocol presented above may be omitted.

\subsection{A simple individual attack}

Without   loss  of   generality,   suppose  Alice   sends  the   state
$|\Phi_+\rangle$ and Eve attacks  fraction $f$ of particles from Alice
to Bob.   Eve has an equal chance  of measuring in the  right or wrong
basis.   If she  measures  in $G_2^A$  (with  probability $f/2$),  she
always obtains  $|\Phi_+\rangle$, which  she forwards to  Bob, without
introducing any error.

On the other hand, if she  measures in a basis other than $G_2^A$, she
finds any one of the  four basis elements with equal probability.  She
forwards the obtained state to Bob.  After Alice's public announcement
of basis, she is equally unsure of what state Alice prepared as she is
of what state  Bob obtained. The error rate $e$  generated is given by
the probability that Alice and  Bob, measuring in the same basis, find
the wrong  outcome, which is:
\begin{equation}
e = \frac{f}{2} \times \frac{3}{4} = \frac{3f}{8}
\label{eq:e}
\end{equation}
Eve's average information (symmetrically with respect to Alice or Bob)
per transmitted  particle is maximal when,  during Alice's announcement
of bases, she finds that  Alice's basis matches hers, and minimal when
it does not. In terms of error observed, her information is:
\begin{equation}
I(A:E) =  I(B:E)   =   2 \times \left(\frac{f}{2}\right)
 = \frac{8e}{3} ~\textrm{bits},
\label{eq:IAE}
\end{equation}
in  view of  Eq.  (\ref{eq:e}).  Because  of  the mutual  unbiasedness
property between  any two bases,  after Alice and Bob  have reconciled
their bases, Eve's action induces  on any input symbol $m$, the output
probability distribution $P(n)$ where $P(n=m)  = 1-e$ and $P(n\ne m) =
\frac{e}{3}$.  The  corresponding Shannon entropy  functional is given
by   $H\left(1-e,  \frac{e}{3},   \frac{e}{3},   \frac{e}{3}\right)  =
-(1-e)\log_2(1-e) -e\log(e/3)$.

\begin{figure}
\includegraphics[width=14.0cm]{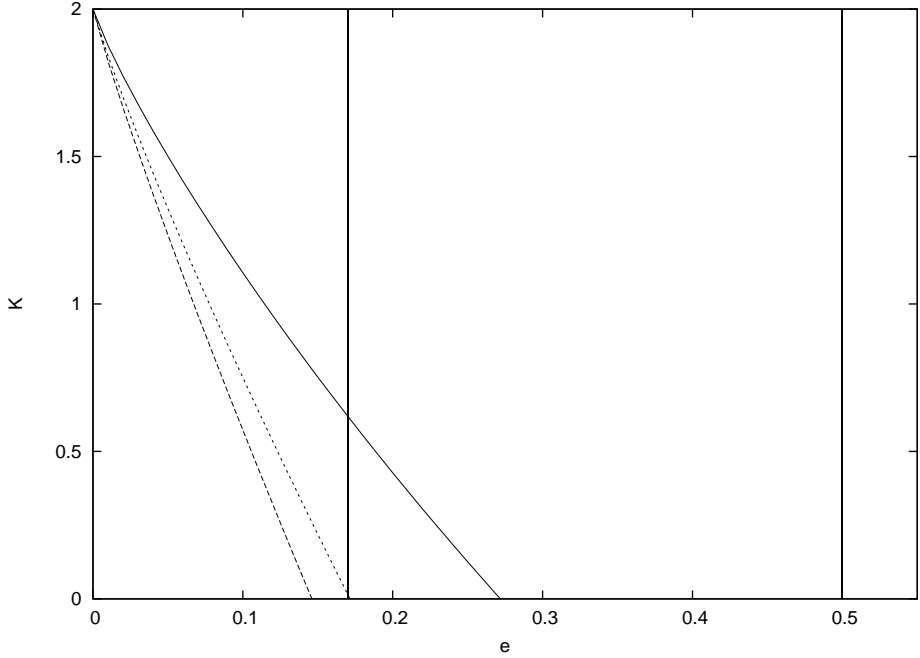}   
\caption{\bla The secret  key rate as a function of  error rate $e$ in
  the  conventional  attack  scenario (Eq.   (\ref{eq:IAB})  and,  as
  explained   later,    the   side-channel   attack    scenario   (Eq.
  (\ref{eq:IABaprime}).   The   conventional  security   scenario  is
  represented by  the rightmost curve,  which has a positive  key rate
  while  $e \le e_2 \approx 27\%$, the largest tolerable error
  rate in the simple individual attack in a conventional cryptographic
  scenario.  The leftmost curve  represents the above attack augmented
  in a side-channel scenario, with  Eve's maximal attack ($F=\half$ in
  Eq.  (\ref{eq:newIABprime})), with a tolerable error rate of at most
  14.5\% for depolarizing  action Eq.  (\ref{eq:wernere}), the
  noisy  state is  nonlocal for  $e <  e_{\rm LR}  \approx 0.17$  (Eq.
  (\ref{eq:eloc}), first vertical line) and  entangled for $e < e_{\rm
    ent} =  0.5$ (Eq.   (\ref{eq:eent}), second vertical  line).  Thus
  the  individual attack  in the  conventional scenario  allows secure
  states that  are local,  but which  are necessarily  entangled.  The
  intermediate  curve  represents  $F=0.6$  for which  the  region  of
  nonlocality and secrecy coincide.}
\label{fig:ir}
\end{figure}

Assuming  Alice  sends   all  four states  in  both   bases  with  equal
probability, Bob's information is given by the mutual information:
\begin{equation}
I(A:B) = 2 - H\left(1-e,\frac{e}{3},\frac{e}{3},\frac{e}{3}\right).
\label{eq:IAB}
\end{equation}
The condition for a positive key rate is
\begin{equation}
K = I(A:B) - \min\{I(A:E), I(B:E)\} > 0,
\label{eq:poskey}
\end{equation}
whose sign  is determined  by Eqs.  (\ref{eq:IAB})  and (\ref{eq:IAE})
\cite{ck78}.  $K$ is  the secret bit rate that can  be distilled.  The
key  rate for  this situation  is plotted  as the  rightmost curve  in
Fig. \ref{fig:ir}.   The largest tolerable error  rate, which we
denote by $e_2$, is about  27 
the relative  weakness of Eve's  attack. As  we find later,  even this
weak attack, augmented by Eve's access to certain side channels, leads
to more stringent bounds (the other two curves in the same Figure).

By Eve's  interference, she is  acting as a depolarizing  channel that
has the action:
\begin{equation}
\rho_j^k        \longrightarrow         \mathcal{E}(\rho_j^k)         =
\left(1-\frac{f}{2}\right)\rho_j^k                                    +
\frac{f}{2}\frac{I_4}{4}
\label{eq:werner}
\end{equation}
where $k  \in \{1,2\}$  labels the basis  and $j$  $(\in \{1,2,3,4\})$
labels  the basis  elements.  If  Alice transmits  an  entangled basis
element,  then the  state in  Eq.  (\ref{eq:werner})  is (up  to local
unitaries) a Werner state \cite{werner}.

Substituting   for   $f$  from   Eq.    (\ref{eq:e})  in   Eq.
(\ref{eq:werner}), we find Eve's depolarizing channel in terms of
error rate:
\begin{equation}
\rho_j^k         \longrightarrow        \mathcal{E}(\rho_j^k)        =
\left(1-\frac{4e}{3}\right)\rho_j^k                                   +
\frac{4e}{3}\frac{I_4}{4}, 
\label{eq:wernere}
\end{equation}

\subsection{Entanglement considerations}

Let the measurement settings on the  first and second qubit (i.e., the
polarization and path qubit) be given by the following directions:
\begin{eqnarray}
\vec{a}_{1}=\hat{i}, \vec{a}_{2}=\hat{j},
\vec{a}_{3}=\frac{1}{\sqrt{2}}\hat{i}+\frac{1}{\sqrt{2}}\hat{j},\nonumber\\
\vec{b}_{1}=\frac{1}{\sqrt{2}}\hat{i}+\frac{1}{\sqrt{2}}\hat{j},
\vec{b}_{2}=\frac{-1}{\sqrt{2}}\hat{i}+\frac{1}{\sqrt{2}}\hat{j},
\vec{b}_{3}=\hat{j}
\label{eq:settings}
\end{eqnarray}
\bla which are used for evaluating the following Bell inequality
\begin{equation}
S  =  E(\vec{a}_{1},  \vec{b}_{1})  +  E(\vec{a}_{2},  \vec{b}_{1})  +
E(\vec{a}_{1}, \vec{b}_{2}) -E(\vec{a}_{2},\vec{b}_{2}),
\label{eq:bellop}
\end{equation}
where $E(\vec{a},\vec{b})$ is expectation  value of measuring the spin
in  the  directions $\vec{a}$  and  $\vec{b}$  in  the two  particles,
respectively.  It is well known  that for local-realist models, $S \le
S_{LR} = 2$.  The correlation for the singlet is given by
\begin{eqnarray}
E(\vec{a}_{i},\vec{b}_{j})=-\vec{a}_{i}\cdot\vec{b}_{j},
\label{eq:corr}
\end{eqnarray}
so that $S = -2\sqrt{2} = \sqrt{2}S_{LR}$.
The most general separable state is given by
\begin{eqnarray}
\rho_{sep}=\int\int\sigma(\vec{n}_{a},\vec{n}_{b})|n_{a}\rangle\langle
n_{a}|\otimes |n_{b}\rangle\langle n_{b}|d\vec{n}_{a}d\vec{n}_{b},
\label{eq:rhosep}
\end{eqnarray}
where
$\int\int\sigma(\vec{n}_{a},\vec{n}_{b})d\vec{n}_{a}d\vec{n}_{b}=1$
and
\begin{eqnarray}
\vec{n}_{a}        &=&       \sin\theta_{a}\cos\phi_{a}\hat{i}       +
\sin\theta_{a}\sin\phi_{a}\hat{j}+
\cos\theta_{a}\hat{k}{}\nonumber\\           \vec{n}_{b}           &=&
\sin\theta_{b}\cos\phi_{b}\hat{i}+\sin\theta_{b}\sin\phi_{b}\hat{j}+
\cos\theta_{b}\hat{k}
\end{eqnarray}
The correlations for $\rho_{\sep}$ can be calculated as
\begin{eqnarray}
E(\vec{a}_{i},\vec{b}_{j})= \textrm{Tr}[\rho_{\sep}
\vec{\sigma}.\vec{a}_{i}\otimes\vec{\sigma}.\vec{b}_{j}].
\end{eqnarray}
Using   equations    Eqs.    (\ref{eq:settings}),   (\ref{eq:bellop}),
(\ref{eq:corr}) and  (\ref{eq:rhosep}), the  upper and lower  bound of
the Bell quantity $S$ in Eq. (\ref{eq:bellop}) is given by
\begin{eqnarray}
S &=&\sqrt{2}\int\int\int\int\sigma(\theta_{a},\theta_{b},\phi_{a},\phi_{b})
\sin^{2}\theta_{a}\sin^{2}\theta_{b}\sin(\phi_{a}+\phi_{b})
d\theta_{a}d\theta_{b}d\phi_{a}d\phi_{b}{}\nonumber\\
&\Rightarrow& -\sqrt{2}\leq S \leq \sqrt{2},
\label{eq:bellbound}
\end{eqnarray}
so  that  quantum  bound  for separable  states,  $S_{\rm  max:sep}  =
\sqrt{2} < S_{LR}$ \cite{roy04}.

State $|\Phi^+\rangle$  is equivalent upto  local unitaries to  a Bell
state, and yields the maximal Bell-inequality violation of $2\sqrt{2}$
for settings (\ref{eq:settings}), whereas $S(I_4) = 0$.  After Alice's
public announcement of bases, if  Bob divides the received states into
sub-ensembles corresponding to  each input state, then in  view of Eq.
(\ref{eq:wernere}),   for   the   $|\Phi^+\rangle$  subensemble,   Bob
will observe:
\begin{equation}
S = \left(1 - \frac{4e}{3}\right)2\sqrt{2}.
\label{eq:vio}
\end{equation}
This is nonlocal when $\langle S\rangle > S_{LR} = 2$, or
\begin{equation}
e  <  \frac{3}{4}\left(1  -  \frac{1}{\sqrt{2}}\right)  \equiv  e_{LR}
 \approx 22.5\%.
\label{eq:eloc}
\end{equation}
From  Eq. (\ref{eq:wernere}),  setting  $\left(1-\frac{4e}{3}\right) >
\frac{1}{3}$   as   the  necessary   and   sufficient  condition   for
entanglement  \cite{werner} of  Werner states  by  the Peres-Horodecki
positive-partial-transpose  criterion \cite{ppt},  we find  that Bob's
states are entangled when
\begin{equation}
e < \frac{1}{2} \equiv e_{\rm ent}.
\label{eq:eent}
\end{equation}
Since $e_2  > e_{LR}$ (cf. Fig. \ref{fig:ir}),  it follows
that there  are local  states that allow  secrecy extraction  for both
protocols under the considered attack.  This will not be true when the
side-channel attack is included.

The  corresponding values  of $S$  are, from  Eq.  (\ref{eq:wernere}),
$S_2  = \left(1  -  \frac{4e_2}{3}\right)2\sqrt{2} \approx  1.36S_{\rm
  max:sep}$,  implying   that  $e_2   <  e_{\rm  ent}$   (cf.   Figure
\ref{fig:ir}).  In  other words, all states secure  under the protocol
for the given  class of attacks are necessarily  entangled.  Since the
considered  attacks are  clearly not  the strongest  possible  for the
protocol  considered,  this implies  that  security  or  secrecy is  a
strictly stronger  condition than entanglement (in  that more powerful
attacks will  reduce the tolerable  error rate, and thus  increase the
amount entanglement in the state at the security threshold).

Our above  results may be  compared and contrasted  with corresponding
results  obtained in  the  inter-particle case  for  the link  between
nonlocality and secrecy in quantum mechanics and general non-signaling
theories  in the  conventional attack  and attack  scenarios involving
untrusted  devices   \cite{BHK05,scagis1,scagis2,mag06,AGM06,pir09}. 

\section{Side-channel attacks and faulty devices \label{sec:sidecha}}

The  peculiar  nature of  intra-particle  entanglement  is that  simple
operations like application of an optical element on an arm can be an
entangling operation.  This property  can be useful to protect secrecy
in a side scenario.  As an example, consider  the tiny angular momentum
acquired by the  QWP through recoil during its  rotation of the photon
polarization. Eve may be able to somehow monitor the vibrational state
of the QWP and deduce  private information about the settings used by
Alice.  Alternatively,  Eve may  detect a gap  in the wall  of Alice's
station, and shine a thin pencil  of light beam at some of the optical
elements through the gap and  deduce information based on the pattern
of electromagnetic scatter.   Worse still, Eve may be  the vendor from
whom  Alice and  Bob purchase  their  optical elements.   Even if  the
available  side channels  are  weak, Eve  may  install hidden  "trojan
horses" that reveal basis or outcome information to her.

The  attack implemented  in a  QWP can  be mathematically  modeled as
follows:
\begin{eqnarray}
|b\rangle|A\rangle_D                     &\rightarrow&
|b\rangle|A\rangle_D                          ~(b=0,1)
 \nonumber\\             |0\rangle|P\rangle_D|0_D\rangle_\varphi           &\rightarrow&
|{+}\rangle|P_+\rangle_D|0_D\rangle_\varphi     \rightarrow
|{+}\rangle|P\rangle_D|Y_D\rangle_\varphi \nonumber \\ 
|1\rangle|P\rangle_D|0_D\rangle_\varphi  
 &\rightarrow&           |{-}_D\rangle|P_-\rangle_D|0_D\rangle_\varphi            \rightarrow
|{-}\rangle|P\rangle_D|{Y}_D\rangle_\varphi
\label{eq:scha}
\end{eqnarray}
where  $|A\rangle_D,  |P\rangle_D$  correspond to  states  of the initial
absence or presence of some device (here  the QWP) in path $D \in \{R,
T\}$; $|P_\pm\rangle_D$, the recoiled state  of the device, carrying a
small amount of  angular momentum acquired when the photon  in a $V/H$
state  is  transformed into  one  of   the  diagonal  polarization
states~  $|{\pm}\rangle \equiv  \frac{1}{\sqrt{2}} (|H\rangle  \pm
|V\rangle)$; $|0_D\rangle_\varphi, |Y_D\rangle_\varphi$ are the vacuum
state and state of the  electromagnetic leaking channel, produced when
the device relaxes  back from $|P_\pm\rangle_D$ to  its initial state.
The subscript  in state $|Y_D\rangle$  indicates a photon in  the mode
coupled with device $D$.

Practically speaking,  for Alice to rule out  every possible malicious
defect  is  not  easy,  if  not  impossible.   What  is  desirable  is
statistical  tests on performance  that acts  like a  catch-all check.
The usefulness here of  nonlocal correlations (or, entanglement within
quantum  mechanics) between  measurements by  Alice and  Bob  has been
recognized \cite{BHK05,mag06,AGM06}.   What is interesting  is that
intra-particle entanglement also can be useful in this way.

This is at  first not obvious.  To obtain  the Alice \textendash Bob correlations,
one  of an entangled  pair of  particles must  be transmitted  to Bob,
following   which  each   particle  is   measured   separately.   With
intra-particle entanglement, such  spatial separation is not possible.
Moreover, when Alice transmits  the particle, all entangled degrees of
freedom in  principle become  available to Eve,  who may thus  be able
remove  any trace  of  her tampering  the  devices. In  this work,  we
suggest  that this problem  can be  solved by  having Alice  perform a
Bell-inequality test  on the  particle \textit{after} and all  her devices
have  been  used   (beam  splitters,  polarizers,  measurements),  and
\textit{just before} the particle's transmission to Bob.  Furthermore,
because both  entangled degrees of  freedom are with Alice,  she might
equally  well  measure them  in  $G_2^A$  basis  to verify  their  BIV
property.  This  is accomplished  simply be reversing  the preparation
procedure  and verifying  that the  output is  indeed the  input state
$|V\rangle|\psi_T\rangle$.

The  attack  (\ref{eq:scha})  does  not  affect  elements  from  basis
$G_1^A$,  because no  QWP is  made  use of.   On the  other hand,  any
element  from the  entangled basis  $G_2^A$, say  $|\Phi^+\rangle$, is
affected.  The below analysis holds for any other element in the basis
as well. 

To  prepare the state  $|\Phi_+\rangle \in  G_2^A$, Alice  inputs $|0,
\psi_T\rangle$ into the beam splitter,  and applies a HWP plate on the
$R$  arm to obtain  $\frac{1}{\sqrt{2}}\left(|0\rangle|\psi_T\rangle +
i|1\rangle|\psi_R\rangle\right)$. Applying a QWP  on both arms $R$ and
$T$, in view of Eq.  (\ref{eq:scha}), she effects the transformation:
\begin{eqnarray}
|0\rangle|\psi_T\rangle|P,P\rangle_{R,T}|0,0\rangle_\varphi
 &\rightarrow& 
\frac{1}{\sqrt{2}}\left(|0\rangle|\psi_T\rangle                       +
i|1\rangle|\psi_R\rangle\right)  |P,P\rangle_{R,T}  |0,0\rangle_\varphi
\nonumber \\  &
\rightarrow                                                           &
\frac{1}{\sqrt{2}}\left(\frac{|0\rangle_{s}+|1\rangle_{s}}{\sqrt{2}}
|0_R,Y_T\rangle_\varphi |\psi_{T}\rangle_{p} + i \frac{|0\rangle_{s}
  - |1\rangle_{s}}{\sqrt{2}}|{Y}_R,0_T\rangle_\varphi
|\psi_{R}\rangle_{p}\right)|P,P\rangle_{R,T} \nonumber  \\  &   \ne  &
|\Phi_+\rangle|x,y\rangle_{R,T}|0,0\rangle_\varphi, \label{eq:atak}
\end{eqnarray}
for any $|x,y\rangle$.  Eve needs to distinguish between the case that
the mode $\varphi$  remains vacuum, and that there is  a radiation of a
photon.

Let  the  projector  to  $|0_R,0_T\rangle$ be  denoted  $\Pi_0$.   Let
$\langle  0|Y\rangle   \equiv  \cos\theta$  in  both   arms,  so  that
${_\varphi\langle} Y_R,0_T|0_R,{Y}_T\rangle_\varphi = \cos^2(\theta)$.
It  is easy  to verify  that the  probability of  observed  error upon
Alice's reversing her state preparation is
\begin{equation}
e_A = \frac{1}{2}\sin^2(\theta).
\label{eq:perror}
\end{equation}
This is the probability that, upon reversing her preparation procedure
on a  given particle,  Alice fails  to find it  in the  original input
state  $|0\rangle|\psi_T\rangle$.   We  call this  the  \textit{device
  error}, to contrast  it with the conventional error,  $e$, which may
be called the \textit{channel error}, acquired during the transmission
of  the particle through  the channel.  We will  conservatively assume
that no new noise above the preparation noise is introduced during the
reversing step. 

The reversal is assumed to be implemented through an auxiliary optical
set-up. For quantum cryptography to be a realistic enterprise, we must
assume that  Eve's side channel is \textit{passive},  i.e., she cannot
use  this side  information to  alter any  settings of  Alice's device
(otherwise--  i.e., if  Eve had  \textit{active} access \textemdash clearly Eve
would be too powerful for  security to be meaningful).  In particular,
Eve cannot alter Alice's readings  in this check. Moreover, Eve cannot
access  the  random number  generator  Alice  used  to prepare  random
states. Thus a side-channel  attack of the type (\ref{eq:scha}) cannot
be correlated with  the prepared states, and Eve  cannot hope to lower
the value  of $e_A$ as  seen by  Alice. At best,  Eve can find  out the
value  of  $e_A$,  but  ideally  she knows  this  already,  being  the
adversary who causes the device noise.

We note that this error check  is a simple substitute for a Bell test,
in that  it effects  a measurement that  verifies that  the particle's
state is indeed $|\Phi_+\rangle$.   By contrast, in the inter-particle
case, local  operations and  classical communication would  be needed,
not to mention the difficulty in preparing the entanglement.

What  is   interesting  here,   in  contrast  to   the  inter-particle
entanglement  case,  is that  the  attack  (\ref{eq:scha}) can  render
intra-particle  separable states  as  intra-particle entangled,  again
disrupting the correlations, which can be detected in the verification
step.  For example, the  attack (\ref{eq:scha}) when Alice attempts to
prepare  the  separable   state  $\frac{1}{2}(|0\rangle  +  |1\rangle)
(|\psi_R\rangle  +  |\psi_T\rangle) \in  G^A_4$  instead produces  the
entangled state
\begin{equation}
\frac{1}{\sqrt{2}}\left(\frac{|0\rangle_{s} + |1\rangle_{s}}{\sqrt{2}}
\right)\otimes \left(|0_R,Y_T\rangle_\Phi |\psi_{T}\rangle_{p} + 
i |{Y}_R,0_T\rangle_\Phi
|\psi_{R}\rangle_{p}\right)|P,P\rangle_{R,T}.
\end{equation}
As before, the  verification step detects Eve  with probability $e_A$.
Thus it  is straightforward to  adapt the  analysis given below  for a
protocol where bases $\{G_1^A, G_4^A\}$  are used instead of $\{G_1^A,
G_2^A\}$.

For  the  present   protocol,  one  strategy  for  Eve   would  be  as
follows. She measures the electromagnetic modes in the basis $\{\Pi_0,
1-\Pi_0\}$,              where              $\Pi_0              \equiv
|0_R,0_T\rangle_\Phi\langle0_R,0_T|$ in the  Hilbert space given by
span$\{|0_R,0_T\rangle,|1_R,0_T\rangle,|0_R,1_T\rangle\}$.          The
outcome   $\Pi_0$   is    indeterminate,   while   outcome   $1-\Pi_0$
deterministically  informs Eve that  the basis  $G_2^A$ was  used, and
leaves the particle in the state
\begin{equation}
|\underline{\Phi}_+\rangle \equiv \frac{(1-\Pi_0)|\Phi_+\rangle}
{||(1-\Pi_0)|\Phi_+\rangle||},
\end{equation}
with probability 
\begin{eqnarray}
 P(1-\Pi_0|G_2^A) &\equiv& \langle \Phi_+|(1-\Pi_0)|\Phi_+\rangle \nonumber
 \\   &=&  1-   ||\cos(\theta)|\Phi_+\rangle||^2   \nonumber  \\   &=&
 1-\cos^2(\theta) \nonumber \\ &=& 2e_A,
\label{eq:pcert}
\end{eqnarray}
where the last follows from  Eq. (\ref{eq:perror}). Thus, the more the
deterministic information Eve acquires,  the larger is the disturbance
she produces, that Alice can see.

It can be  shown that the disturbed versions of  the elements in basis
$G_2^A$ remain orthogonal. For example,
\begin{eqnarray}
\langle \underline{\Phi}_+|\underline{\Phi}_-\rangle
&\propto& \langle \Phi_+|(1 - \Pi_0)^2|\Phi_-\rangle \nonumber \\
&=& \langle \Phi_+|(1 - \Pi_0)|\Phi_-\rangle \nonumber \\
&=& -\langle \Phi_+|\Pi_0|\Phi_-\rangle \nonumber \\
&=& -\cos^2\theta \langle \Phi_+|\Phi_-\rangle
\nonumber \\
&=& 0.
\label{eq:ortho}
\end{eqnarray}
Thus, there  exists  a  projective measurement  strategy  whereby  the
elements   of  the  "disturbed"   basis  $\underline{G}^A_2$   can  be
deterministically distinguished. 

Since Eve's  attack on  any element in  $G_1^A$ produces  no radiative
emission, we have the conditional probability
\begin{equation}
P(\Pi_0|G_1^A) = 1.
\label{eq:cond1}
\end{equation}
The   probability  that   Eve  obtains   outcome  $\Pi_0$   is,  using
Eq. (\ref{eq:pcert})
\begin{eqnarray} 
P(\Pi_0) &=& P(\Pi_0|G_1^A)p_1 + P(\Pi_0|G_2^A)(1-p_1) \nonumber \\
   &=& \cos^2(\theta) + p_1\sin^2(\theta),
\label{eq:ptotal}
\end{eqnarray}
where $p_1  \equiv P(G_1^A)$, the  probability that Alice  chooses (an
element of) $G^A_1$. By the Bayesian rule, using Eqs. (\ref{eq:cond1})
and (\ref{eq:ptotal}), we have
\begin{eqnarray}
P(G_1^A|\Pi_0) &=& \frac{P(\Pi_0, G_1^A)}{P(\Pi_0)} \nonumber \\
    &=& \frac{P(\Pi_0|G_1^A)p_1}{\cos^2(\theta) + p_1\sin^2(\theta)} \nonumber\\
    &=& \frac{p_1}{\cos^2(\theta) + p_1\sin^2(\theta)},
\label{eq:req}
\end{eqnarray}
which we  denote $p_{10}$. It may  be noted that if  $p_{10} > \half$,
i.e.,
\begin{equation}
p_1 > \frac{\cos^2(\theta)}{1+\cos^2(\theta)},
\label{eq:p1}
\end{equation}
which happens when $\theta>0$, then Eve's best guess to minimize error
in an intercept-resend attack would be to assume that $G_1^A$ was used
if  her  measurement  returns $\Pi_0$.  That  is  to  say that  if  no
radiation leakage is found  even with high distinguishability, chances
are that the state sent by Alice was an element in $G_1^A$.

In practice, however,  Alice and Bob, could respond  to this potential
tactic  by Eve  this  by always  using  $G_2^A$.  To  avoid this,  Eve
responds  to $\Pi_0$  outcomes  by transmitting  a  random element  of
either basis  chosen with equal  probability.  Eve's full  strategy in
this attack scenario is that she will use this extra basis information
to improve  her guess work  in the intercept-resend attack  of Sect.
\ref{sec:scha}.  If Eve obtains  outcome $1-\Pi_0$, she determines the
$G_2^A$  element obtained  by a  projective measurement,  and forwards
this state to  Bob, producing no errors.  When  Eve obtains an outcome
$\Pi_0$,  she measures  randomly in  either basis,  notes  the outcome
state and forwards it to Bob.   In this case, she identifies the sent
state correctly  if she  measures in the  right basis: in  the $G_1^A$
case, there is no state distortion; in the $G_2^A$ case, orthogonality
is preserved on account of Eq.  (\ref{eq:ortho}).
 
Therefore, the probability she produces  an error that can be detected
by   Alice  and   Bob  is   found,  using   Eqs.   (\ref{eq:req})  and
(\ref{eq:ptotal}), to be
\begin{eqnarray}
e &=& fP(\Pi_0)\half \times \frac{3}{4} \nonumber \\
      &=& \frac{3}{8}\left(\cos^2\theta + p_1\sin^2\theta\right)
\label{eq:eprime}
\end{eqnarray}
in  place  of  Eq.  (\ref{eq:e}),  the  corresponding  error  in  the
conventional   scenario.   Eve's   information   is,   in   place   of
Eq. (\ref{eq:IAB}), now given by by:
\begin{eqnarray}
I^\prime(A:E) = I^\prime(B:E) &=&  2P(1-\Pi_0)f + P(\Pi_0)f \nonumber \\
&=& \frac{8e}{3}\frac{1 + \sin^2\theta(1-p_1)}{1 - \sin^2\theta(1-p_1)},
\nonumber \\
&=& \frac{8e}{3}\frac{1 + 2(1-F)(1-p_1)}{1 - 2(1-F)(1-p_1)},
\label{eq:IABaprime}
\end{eqnarray}
where 
\begin{equation}
F = 1-e_A,
\label{eq:F}
\end{equation}
where $F$ is  the probability of recovering $|0\rangle|\psi_T\rangle$,
upon  reversing and  measuring in  the verification  step,  and taking
values  between  1  (when   Eve  does  not  attack,  corresponding  to
$\theta=0$) to  $\half$ (when Eve maximally  attacks, corresponding to
$\theta=\pi/2$).   In   the  limit  of   indistinguishability  of  the
radiative   leak   modes   (i.e.,   $\theta  \rightarrow   0$),   Eqs.
(\ref{eq:eprime})  and (\ref{eq:IABaprime})  reduce,  respectively, to
Eqs.  (\ref{eq:e}) and (\ref{eq:IAB}).

In particular, letting  $p_1=\half$, from Eq. (\ref{eq:IABaprime}), we
have
\begin{equation}
I^\prime(A:E) = I^\prime(B:E) = \frac{8e}{3}\frac{2-F}{F}.
\label{eq:newIABprime}
\end{equation}
Eve's maximal  attack in this model corresponds  to $F=\half$, whereby
she fully  distinguishes the side channels corresponding  to $G_1$ and
$G_2$ ($\theta = \pi/2$  in Eq. (\ref{eq:perror})).  The corresponding
secret key rate (as a  function of \textit{channel error}) obtained by
substituting  Eqs.  (\ref{eq:newIABprime})  and (\ref{eq:IAB})  in Eq.
(\ref{eq:poskey}) is  plotted in  Fig. \ref{fig:ir} as  the leftmost
curve.  The  highest tolerable error rate, which  is the $x$-intercept
of the curve,  is about 14.5  as seen from
Figure \ref{fig:ir}, this implies  that secure states in this scenario
are necessarily Bell-inequality violating.  This is in contrast to the
security in  the conventional scenario (the rightmost  curve in Fig.
\ref{fig:ir}), where there are  secure states that are Bell-inequality
non-violating ($0.17  \lesssim e  \lesssim 0.27$).  The  central curve
corresponds to $F  \approx 0.6$, for which positivity  of secrecy rate
coincides with the BIV property.

For the  attacked state in  Eq. (\ref{eq:atak}), the amount  of Bell's
inequality  violation  can  be  shown  to  be  $\mathcal{B}(\theta)  =
2\sqrt{2}(1+\cos^2(\theta))$,   so   that   it   follows   from   Eqs.
(\ref{eq:perror})  and   (\ref{eq:F})  that  $\mathcal{B}=2\sqrt{2}F$.
Substituting this in Eq.  (\ref{eq:newIABprime}), and allowing for the
possibility of some errors being due  to noise rather than Eve, we can
bound Eve's information from above  as a function of Bell's inequality
violation observed by Alice to be:
\begin{equation}
I^\prime(A:E) \le  \frac{8e}{3}\frac{4\sqrt{2}-\mathcal{B}}{\mathcal{B}},
\label{eq:newIABprimee}
\end{equation}
This  form may  be  compared with  bounds  on Eve's  information as  a
function of  Bell's inequality violation in  DIQKD with inter-particle
entanglement, except that $\mathcal{B}$  is evaluated by Alice, rather
than Bob and indicates the level  of Alice's device  error, which in
turn indicates the level of channel error that can be tolerated.

\section{Conclusions \label{sec:conclu}}

Our   work   proposes   the   use   of   intra-particle   entanglement
(path-polarization  entanglement of  single photons)  for cryptography
using an  interferometric setup.  Here, unlike  the conventional BB84,
the system  used is  four dimensional and hence one can  have five
mutually unbiased bases for encoding.  We illustrate its usefulness by
pointing out a type of side-channel attack which it is secure against,
but  which  renders  a BB84-like  protocol  insecure.   Intra-particle
entanglement is necessarily checked by  a local Bell test, which leads
to  a decoupling  of device  noise  from channel  noise. The  observed
device noise, derived from the local Bell test, determines the channel
error  that can  be tolerated.   Note  that error  rates and  security
proofs have  been explicitly given  in the  paper. We may  also stress
that,  since the  use  of intra-particle  entanglement  allows one  to
distinguish  between  channel  and   device  errors  which,  in  usual
protocols, are  indistinguishable, this could  be of aid  in assessing
the security of the protocol when actually implemented.

The present  work mainly  highlights the usefulness  of intra-particle
entanglement for  QKD. There are  a number of prospects  for extending
this work. The attack model  on the channel can consider more powerful
adversaries,   for   example,  executing   a   coherent  attack.    The
eavesdropping scenario can be expanded from a single untrusted element
toward higher DI.  Ref.  \cite{BCW+12} considers the
problem posed  by optical  losses on the  channel, which  has remained
beyond  the scope  of  the present  work,  and is  of  interest for  a
practical  application.   Similarly,  the  issue of  composability  of
intra-particle entanglement-based  QKD  may  be  studied  \cite{RK04}.
Finally,    the   problem    of    secure   direction    communication
\cite{q1,q2,q3,q4,q5,q6}  with  intra-particle  entanglement would  be
worth considering.


ASM  and  DH  acknowledge  support  from the  Department   of  Science  and
Technology,  India (DST)  Project  SR/S2/LOP-08/2013, and  RS for  the
DST-supported Project  SR/S2/LOP-02/2012.  DH  also thanks  the Centre
for Science, Kolkata for support.


\end{document}